\documentclass[a4paper,10pt,final]{article}
\usepackage{amsfonts}
\usepackage{amsmath}
\usepackage{amssymb}
\usepackage[dvips]{graphicx}

\DeclareGraphicsExtensions{.bmp,.png,.pdf,.jpg,.eps}

\begin{document}

\title{Modeling amortization systems with vector spaces}
\author{J. S. Ardenghi$^{\dag }$\thanks{
email:\ jsardenghi@gmail.com, fax number:\ +54-291-4595142} \\
$^{\dag }$IFISUR, Departamento de F\'{\i}sica (UNS-CONICET)\\
Avenida Alem 1253, Bah\'{\i}a Blanca, Buenos Aires, Argentina}
\maketitle

\begin{abstract}
Amortization systems are used widely in economy to generate payment
schedules to repaid an initial debt with its interest. We present a
generalization of these amortization systems by introducing the mathematical
formalism of quantum mechanics based on vector spaces. Operators are defined
for debt, amortization, interest and periodic payment and their mean values
are computed in different orthonormal basis. The vector space of the
amortization system will have dimension $M$, where $M$ is the loan maturity
and the vectors will have a $SO(M)$ symmetry, yielding the possibility of
rotating the basis of the vector space while preserving the distance among
vectors. The results obtained are useful to add degrees of freedom to the
usual amortization systems without affecting the interest profits of the
lender while also benefitting the borrower who is able to alter the payment
schedules. Furthermore, using the tensor product of algebras, we introduce
loans entanglement in which two borrowers can correlate the payment
schedules without altering the total repaid.
\end{abstract}

\section{Introduction}

Credits induce the design of maturity profiles to decrease the loan
principal. The different amortization systems applied widely in banking and
finance are based on a set of recurrence relations between the debt, the
interest, the amortization and the periodic installments. Once the payment
schedule is defined, an inherent risk shows up at each period due to the
possible borrower default. In \cite{qcl}, the amortization systems has been
studied from a different point of view. Following the trend of applying the
mathematical methods of quantum mechanics in economics \cite{pio0}, in \cite%
{qcl} it has been shown how to obtain the recurrence relations of the loan
by defining a specific algebra of operators. Rewriting the loans on vector
spaces is analogue to the development of the quantum prisoner's dilemma,
where the set of strategies are considered as unitary operators acting on a
Hilbert space. The high degree of adaptability of quantum mechanics resides
in the mathematical flexibility of vector spaces that allows the possibility
of superposition and entanglement of vectors and these principles have been
applied to model decision making (\cite{bus}, \cite{ae4}, \cite{mog}, \cite%
{yuk},\cite{khr}, \cite{pot}, \cite{bus2}, \cite{ger} and \cite{kas}), where
judgments and decisions can be conceived as indeterministic processes when
subjects give answers in situations of uncertainty, confusion or ambiguity,
in econophysics (\cite{sch}, \cite{bag}, \cite{cho}, \cite{ata}, \cite{baq0}%
, \cite{pio1}, \cite{pio2}, \cite{ahn} and \cite{smolin}), where stock
return distributions are modeled by appropriate quantum forces or where
gauge fields are used to model the market dynamics \cite{baq4} and in
quantum game theory (\cite{mey}, \cite{kla}, \cite{eis}, \cite{du} and \cite%
{mar}) where the player strategies are operators and other general aspects
of the human condition (\cite{ae1}, \cite{bor}, \cite{ae2}, \cite{ae3}, \cite%
{kh} and \cite{atm}). In \cite{qcl}, a time evolution of the loan
configuration is obtained through the algebra of loan operators and the mean
values in these configurations give the respective values for the debt,
amortization, interest and periodic payments. When the loan states are
eigenvectors of the loan operators, the mean values are identical to the
eigenvalues, but a rotation of the orthonormal basis of the vector space
induces new values for the debt, amortization, interest and payments
depending on the rotation angle. The benefits of extending the recurrence
relations of a loan to an algebra of operators is due to the access of new
degrees of freedom associated to the allowed rotations. The procedure is to
define a vector space of dimension $M$, where $M$ is the maturity or the
duration of the repayment of the debt. In the moment of contractual
agreement between the lender and the borrower a debt $d_{0}$ is created that
must be returned in a set of scheduled payments obtained from the
amortization system applied by the lender. The time-ordered values of the
loan (debt, amortization, interest and periodic payment) are obtained as the
sequence of mean values of the respective loan operators ($D$, $A$, $Y$ and $%
Q$). The flexibility of the vector space comes from the fact that we can
construct an infinite number of orthonormal basis in which the operators can
take their mean values. Any orthonormal basis can be obtained from another
by a rotation but these does not change the total amortization and the
lender's profit, computed as the sum amortizations (the trace of $A$) and
the sum of the periodic payments (the trace of $Q$). This rotation is a
manifestation of a $SO(M)$ symmetry of the vector space, where $SO(M)$ is
the special orthogonal group of dimension $M$.

In this work, we explore the concept of entanglement by allowing to
correlate two loans. Entanglement is revealed by a violation of Bell-type
inequalities where correlations are obtained for coincidence measurements 
\cite{bell}. These inequalities have been experimentally verified (\cite{asp}%
, \cite{asp2} and \cite{giust}). An entangled state is a quantum state that
cannot be factorized as a product of states of each vectorial space. This
implies that each quantum system cannot be described independently from the
other and this originates correlation between distant measurements that
cannot influence each other \cite{yin}. Entanglement has been applied
outside quantum physics (\cite{bruza}, \cite{bruza2}, \cite{Ney}) where
there can be non-spatial connections between different conceptual entities,
depending on how much meaning they share. In general, entanglement in
quantum games between players is conceived as a kind of mediated
communication or as a contract between the players. In this work,
entanglement between loans is part of a loan agreement between both lenders
and both borrowers where an entanglement parameter is fixed at the beginning
of the repayments and then each borrower can apply an arbitrary rotation
over its own vectorial space.

This work will starts with a brief explanation of the theoretical framework
introduced in \cite{qcl} and the superposition of classical loans will be
discussed. We will go beyond the classical amortization systems and we will
build tensor product spaces of different loans and we will consider
entanglement loan configurations in such a way to obtain entangled payment
schedules for different borrowers. The results obtained in \cite{qcl} and
the given in this work are useful to study how can loans can be redesigned
to reduce macrovolatity and default instead of designing countercyclical
payments. The large degrees of freedom given by the parameter space of the $%
SO(M)$ symmetry is suitable to tune the maturity profile without altering
the lender profit. This manuscript will be organized as follows:\ In Section
II, indexed credit loans are reviewed. In Section III, the recurrence
relations for the debt, amortization, interest and periodic installments are
described in terms of a Generalized Heisenberg algebra and the superposition
of loan configurations is studied showing that classical amortization
systems can be applied simultaneously. Within the same section, loans
entanglement is presented by defining an entanglement matrix that correlates
the loan configurations and where the borrowers can apply its own rotation.
Finally, the conclusions are presented.

\section{Indexed credit loans}

The amortization systems are financial instruments to be used to repay an
initial debt and the interest. These systems consist on three coupled
recurrence relations between the debt $D$, the amortization $A$, the payment 
$Q$, the interest $Y$ and the interest rate $T$%
\begin{equation}
\text{a) }q_{n}=a_{n}+y_{n}\text{ \ \ b) }y_{n}=t_{n-1}d_{n-1}\text{ \ \ c) }%
d_{n+1}=d_{n}-a_{n+1}  \label{c1}
\end{equation}%
where we have introduced a variable interest rate $t_{n}$ to generalize
Wq.(1) of \cite{qcl}. The boundary conditions are an initial debt obligation 
$d_{0}$ and $d_{M}=0$ where $M$ is the maturity of the loan. Summing in $%
a_{n}$, the total amortization $\sum\limits_{n=1}^{M}a_{n}=d_{0}$ repays the
initial debt. Combining the three equations of Eq.(\ref{c1}) we obtain the
following recurrence relation for $d_{n}$ 
\begin{equation}
d_{n}=(1+t_{n})d_{n-1}-q_{n}  \label{c2}
\end{equation}%
In general, the lender fix $t_{n}=t$, but for indexed loans, for example in
non-monetary units that are linked to the inflation rate or any other
macroeconomical variable, it is possible to obtain $t_{n}$ in terms of the
fixed interest rate $t$. For simplicity we consider $t_{n}=t$ and the
solution of last equation reads%
\begin{equation}
d_{n}=(1+t)^{n}\left[ d_{0}-\sum\limits_{j=1}^{n}\frac{q_{j}}{(1+t)^{j}}%
\right]  \label{c3}
\end{equation}%
By applying the boundary condition $d_{M}=0$ we obtain that the periodic
payments obey the restriction 
\begin{equation}
d_{0}=\sum\limits_{j=1}^{M}\frac{q_{j}}{(1+t)^{j}}  \label{c4}
\end{equation}%
The main difference between the loans lies in the way the payment schedule
is computed:

\begin{itemize}
\item French system (annuity amortization): The periodic payments are
constant $q_{j}=q$ and from Eq. (\ref{c4}) we obtain $q_{F}=\frac{%
d_{0}t(1+t)^{M}}{(1+t)^{M}-1}$. The debt and the amorization in the French
system reads%
\begin{equation}
d_{n}^{(F)}=\frac{d_{0}}{(1+t)^{M}-1}\left[ (1+t)^{M}-(1+t)^{n}\right] \text{
\ \ \ \ \ }a_{n}^{(F)}=(1+t)^{n}(q^{(F)}-td_{0})  \label{c5}
\end{equation}

\item German system:\ The amortization is constant $a_{n}=a_{G}$ and using $%
\sum\limits_{n=1}^{M}a_{n}=d_{0}$ we obtain $\ a_{G}=d_{0}/M$. The debt and
periodic payments read%
\begin{equation}
d_{n}^{(G)}=d_{0}(1-\frac{n}{M})\text{ \ \ \ \ }q_{n}^{(G)}=\frac{d_{0}}{M}%
\left[ 1+t(M-n+1)\right]  \label{c6}
\end{equation}%
From last equation it can be seen that the payments are not constant but
they obey Eq.(\ref{c4}).

\item Interest-only system:\ The interest is constant $y_{n}=y_{A}=td_{0}$,
the debt is $d_{n}^{(A)}=d_{0}$ for $n=1,...,M-1$ and the amortization and
periodic payments read%
\begin{gather}
a_{n}^{(A)}=0\text{\ \ \ \ \ \ }a_{M}^{(A)}=d_{0}  \label{c7} \\
q_{n}=td_{0}\text{\ \ \ \ \ \ }q_{M}=(1+t)d_{0}  \notag
\end{gather}

\item Bullet loan system (negative amortization): The loan has to be repaid
at maturity and the only non-zero periodic payment is the last one, then $%
q_{n}^{(B)}=0$ for $n=1,...,M-1$ and $q_{M}=d_{0}(1+t)^{M}$. The debt and
amortization reads%
\begin{gather}
d_{n}^{(B)}=(1+t)^{n}d_{0}\text{ \ \ }\ \ \ \,d_{M}^{(B)}=0  \label{c8} \\
a_{n}=-t(1+t)^{n-1}d_{0}\text{ \ \ \ \ \ \ }a_{M}=(1+t)^{M-1}d_{0}  \notag
\end{gather}
\end{itemize}

These are the most commonly used loan repayment schedules and interest
calculation techniques (for more details see \cite{loan1}). These
amortization systems obey Eq.(\ref{c4}) which gives a restriction over the
possible values of $q_{j}$. This equation is suitable to relax the payment
schedule, for example in the French system. We can demand that $%
Q=\sum\limits_{j=1}^{M}q_{j}=q_{F}M$, which implies that the total amount
paid $Q$ by the borrower with non-constant payments is identical to the
total amount paid in the French system. For instance, with $M=3$ we have
that $Q=3q_{F}=\frac{3D_{0}t(1+t)^{3}}{3t+3t^{2}+t^{3}}$ then we have two
equations to be obeyed by $q_{1}$, $q_{2}$ and $q_{3}$%
\begin{gather}
q_{1}+q_{2}+q_{3}=\frac{3d_{0}t(1+t)^{3}}{3t+3t^{2}+t^{3}}  \label{c9} \\
\ d_{0}=\frac{q_{1}}{1+t}+\frac{q_{2}}{(1+t)^{2}}+\frac{q_{3}}{(1+t)^{3}} 
\notag
\end{gather}%
By solving for $q_{2}$ and $q_{3}$ as a function of $q_{1}$ we obtain%
\begin{gather}
q_{1}=q_{1}  \label{c10} \\
q_{2}=\frac{d_{0}}{3+t(3+t)}\left[ (1+t)^{3}(3+t)-q_{1}(2+t)(3+t(3+t))\right]
\notag \\
q_{3}=q_{1}(1+t)-\frac{d_{0}t(1+t)^{3}}{3+t(3+t)}  \notag
\end{gather}%
The $q_{1}$ value is arbitrary although it must satisfy two restrictions.
The first is given by $q_{2}\geq 0$ which implies that $q_{1}\leq \frac{%
(1+t)^{3}(3+t)}{(2+t)(3+t(3+t))}$ and the second restriction comes from $%
q_{3}\geq 0$ which implies $q_{1}\geq \frac{d_{0}t(1+t)^{2}}{3+t(3+t)}$. 
\begin{figure}[tbp]
\centering\includegraphics[width=98mm,height=62mm]{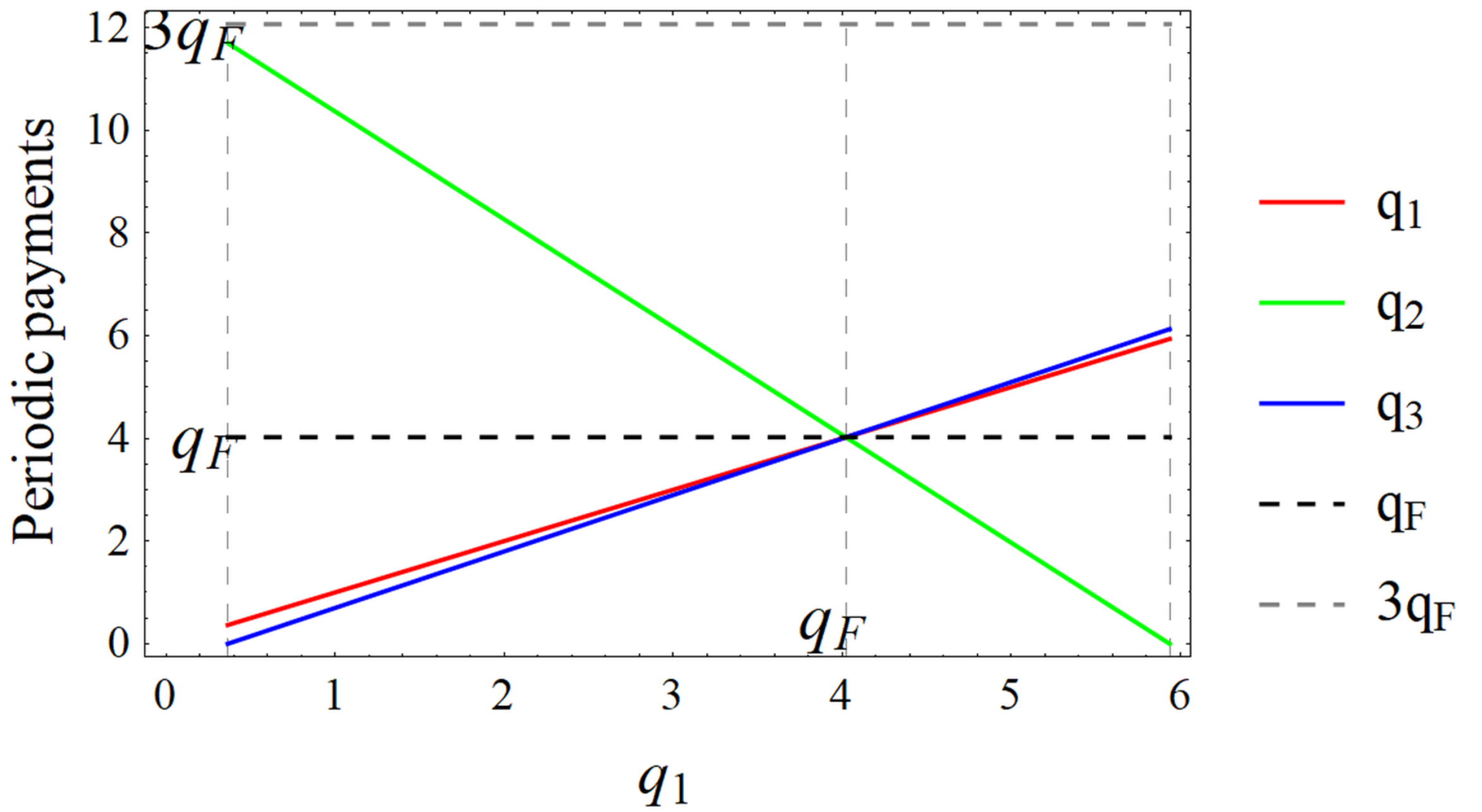}
\caption{Flexible French system with $M=3$, $d_{0}=100$ y $t=0.2$. Different
payment schedules are obtained as a function of the initial payment $q_{1}$.
When $q_{1}=q_{F}$ all the payment schedules are identical to $q_{F}$. }
\label{franfle}
\end{figure}
In Fig. \ref{franfle}, it can be seen $q_{1}$, $q_{2}$ y $q_{3}$ as a
function of $q_{1}\,$and the constant periodic payment $q_{F}$ of the usual
French system and the total amount paid $\mathbf{Q}_{F}=3q_{F}$. In the
figure it can be seen the different payment schedules as a function of $%
q_{1} $ that differ from the usual French system. Two extreme cases can be
obtained when $q_{2}=0$ or $q_{3}=0$ in which one of the periodic payments
is zero but this gives a loan with a $M=2$. In turn, from Fig. \ref{franfle}
we can note that when $q_{1}<q_{F}$, then $q_{2}>q_{F}$ and $q_{3}<q_{F}$
and when $q_{1}>q_{F}$, then $q_{2}<q_{F}$ and $q_{3}>q_{F}$. This indicates
that we can group\ the possible payment schedules for $M=3$ with alternating
payments around the French system payment. Should be noted that the usual
French system cannot be adapted to the German system because there is no
payment schedule where all the installments decrease in time and the total
amount paid is $q_{F}M$. The constraint of Eq.(\ref{c4}) reduces the
possible values of the payments $q_{i}$ but allows different payment
configurations with equal sum. This simple example shows that even the
recurrence relations given by Eq.(\ref{c1}) allows the payment schedule to
be more flexible without altering what the lender earns.

\section{Amortization systems over vectorial spaces}

In \cite{qcl} the amortization systems have been generalized by obtaining
the coupled recurrence relations of Eq.(\ref{c1}) as relations between
eigenvalues of operators acting on a vectorial space. We have introduced a
set of operators $D$, $A$, $Q$, $Y$ and $T$ whose eigenvalues are the debt $%
d_{n}$, the amortization $a_{n}$, periodic payment $q_{n}$, interest $y_{n}$
and interest rate $t_{n}$, respectively. The eigenvalue-eigenvector
equations for these operators read%
\begin{gather}
D\left\vert n\right\rangle =d_{n}\left\vert n\right\rangle \text{ \ \ \ }%
A\left\vert n\right\rangle =a_{n}\left\vert n\right\rangle  \label{ex1} \\
Y\left\vert n\right\rangle =y_{n}\left\vert n\right\rangle \text{ \ \ \ \ \ }%
Q\left\vert n\right\rangle =q_{n}\left\vert n\right\rangle  \notag
\end{gather}%
where $\left\vert n\right\rangle $ are the simultaneous eigenvectors, which
we will call loan configurations, of $D$, $A$, $Y$, $Q$ and $T$ operators.
Last equations imply that%
\begin{equation}
\left[ D,A\right] =\left[ D,Y\right] =\left[ D,Q\right] =\left[ A,Y\right] =%
\left[ A,Q\right] =\left[ Y,Q\right] =0  \label{ex2}
\end{equation}%
which means that all the loan operators commute each other and there is no
incompatibility between them. For simplicity we can assign a debt,
amortization, interest and periodic payment values at each period by
computing the mean value of the respective loan operators in the loan
configuration $\left\vert n\right\rangle $%
\begin{gather}
\left\langle D\right\rangle _{n}=\left\langle n\right\vert D\left\vert
n\right\rangle \text{ \ \ \ \ \ \ }\left\langle Y\right\rangle
_{n}=\left\langle n\right\vert Y\left\vert n\right\rangle  \label{ex3} \\
\text{ \ }\left\langle A\right\rangle _{n}=\left\langle n\right\vert
A\left\vert n\right\rangle \text{ \ \ \ \ \ \ \ }\left\langle Q\right\rangle
_{n}=\left\langle n\right\vert Q\left\vert n\right\rangle  \notag
\end{gather}%
In this case the mean values are identical to the eigenvalues because the
loan configurations $\left\vert n\right\rangle $ are eigenvectors of the
loan operators. Recurrence relations can be obtained from a generalized
Heisenberg algebra (GHA), where eigenvalues can be obtained from a
recurrence relation derived from the algebra (\cite{souza} and \cite{curado}%
). A simple derivation of a GHA is by considering three operators $H$, $a$
and $a^{\dagger }$, where $H$ is the Hamiltonian with eigenvectors $%
\left\vert n\right\rangle $ and eigenvalues $\epsilon _{n}$ such that $%
H\left\vert n\right\rangle =\epsilon _{n}\left\vert n\right\rangle $. The
operators $a$ and $a^{\dagger }$ are the annihilation and creation operators
that acts on the Hamiltonian eigenvectors as $a\left\vert n\right\rangle
=N_{n}\left\vert n-1\right\rangle $ and $a^{\dagger }\left\vert
n\right\rangle =N_{n+1}\left\vert n+1\right\rangle $, where $N_{n}$ are
normalization factors. The GHA\ is obtained by considering the following
relations%
\begin{gather}
\text{a) }aH=f(H)a\text{ \ \ \ \ b) }Ha^{\dagger }=a^{\dagger }f(H)
\label{ex4} \\
\text{ \ \ \ c) }\left[ a,a^{\dagger }\right] =f(H)-H  \notag
\end{gather}%
where $f(H)$ is some analytical function of $H$. Using eq.(a) or eq.(b) we
obtain that $\epsilon _{n}=f(\epsilon _{n-1})$ where we have used that $%
f(H)\left\vert n\right\rangle =f(\epsilon _{n})\left\vert n\right\rangle $.
Eq.(c) gives a recurrence relation for the coefficients $N_{n}$ as $%
N_{n+1}^{2}=N_{n}^{2}+f(\epsilon _{n})-\epsilon _{n}$. Given a function $%
f(x) $, $\epsilon _{n}=f(\epsilon _{n-1})$ gives a recurrence relation for
the Hamiltonian eigenvalues $\epsilon _{n}$ similar to those found in the
amortization systems (see Eq.(\ref{c1})). That is, the generalized
Heisenberg algebra restricts the possible values of $\epsilon _{n}$ to those
that obey $\epsilon _{n}=f(\epsilon _{n-1})$, which is identical to the
restrictions imposed in Eq.(\ref{c1}) to the loan values. Then it is
possible to define an analogous algebra for the loan operators. Due to the
simultaneous magnitudes ($D$, $A$, $Y$, $Q$ and $T$) with defined values,
the generalization to several commuting operators of the GHA is
straightforward. Due to the finite loan duration $M$, the algebra must be
defined over a finite-dimensional Hilbert space, in contrast to the algebras
defined in \cite{curado}, where an infinite dimensional Hilbert space is
considered.\footnote{%
Finite dimensional Hilbert spaces has been widely used to model the stock
market that are isomorphic to $%
\mathbb{C}
^{d}$, where $d$ is the discrete number of possible rates of return \cite%
{cotf}.} In \cite{buzek}, a suitable procedure to obtain finite dimensional
Heisenberg algebras is explained which allows us to define the algebra of
loan operators 
\begin{gather}
\text{a) }\left[ D,A\right] =\left[ D,Y\right] =\left[ D,Q\right] =\left[ A,Y%
\right] =\left[ A,Q\right] =\left[ Y,Q\right] =0\text{\ }  \label{new1} \\
\text{b) }aY=TDa\text{\ \ \ \ c)}Ya^{\dag }=a^{\dag }TD  \notag \\
\text{d) }\left[ D,a\right] =aA~\ \ \ \ \text{e) }\left[ a^{\dag },D\right]
=Aa^{\dag }  \notag \\
\text{f)}\ Q=Y+A\ \ \text{g) }\left[ a,a^{\dag }\right] =A-d_{0}\left\vert
M\right\rangle \left\langle M\right\vert  \notag
\end{gather}%
From Eq.(\ref{new1}a)) a set of $M$ orthogonal loan configurations $%
\left\vert n\right\rangle $ exist that are simultaneous eigenstates of $D$, $%
A$, $Y$, $Q$ and $T$ and where $d_{n}$, $a_{n}$, $y_{n},$ $q_{n}$ and $t_{n}$
are the respective eigenvalues. The annihilation operator $a$ and creation
operator $a^{\dag }$ acts on the $\left\vert n\right\rangle $ basis as%
\begin{eqnarray}
a\left\vert 1\right\rangle &=&0\text{ \ \ \ \ \ \ \ \ \ \ }a\left\vert
n\right\rangle =N_{n}\left\vert n-1\right\rangle  \label{new2} \\
a^{\dag }\left\vert M\right\rangle &=&0\text{ \ \ \ \ \ \ \ \ \ }a^{\dag
}\left\vert n\right\rangle =N_{n+1}\left\vert n+1\right\rangle  \notag
\end{eqnarray}%
where $N_{n}$ are normalization factors of the loan GHA and $\left\vert
M\right\rangle $ is the highest level at which the debt eigenvalue is%
\begin{equation}
D\left\vert M\right\rangle =0  \label{new3}
\end{equation}%
that is, the highest debt level eigenvalue is $d_{M}=0$, which implies that
the loan has finished and $M$ is the loan duration. To see how the algebra
works, Eq.(\ref{new1}) b) can be applied to a loan configuration $\left\vert
n\right\rangle $ obtaining $y_{n}=t_{n-1}d_{n-1}$ which is Eq.(\ref{c1})b).
Similarly, Eqs.(\ref{new1}) d) and f) gives the relation between the
amortization, debt, periodic payment and interest (eq.(\ref{c1}) a) and Eq.(%
\ref{c1}) c)). Equations (\ref{new1}) c) and (\ref{new1}) f) are the
Hermitian conjugate of Eqs.(\ref{new1}) b) ad e). Finally, Eq.(\ref{new1})
g)\ defines the commutation relation between $a$ and $a^{\dag }$ in terms of
the amortization operator $A$. This equation implies that the total
amortization repays the initial debt%
\begin{equation}
Tr([a,a^{\dagger }])=Tr(A)-d_{0}Tr(\left\vert M\right\rangle \left\langle
M\right\vert )=0  \label{new4}
\end{equation}%
Once the eigenvalues of the loan operators are related through the
recurrence relations of Eq.(\ref{c1}), the temporal evolution of \ the
payments must be described. The discrete index $n$ can be used as the loan
time evolution by selecting one by one the unit vectors of the eigenbasis
that diagonalizes simultaneously the loan operators. The increasing value of 
$n$ can be obtained by applying the creation operator $a^{\dagger }$
successively to the ground loan configuration $\left\vert 1\right\rangle $.
This implies that we can obtain the time evolution of the amortization
system by evolving the loan operators as $a^{n}O(a^{\dag })^{n}$ where $O$
can be any loan operator.

\subsection{Superposition}

The main advantage of vectorial spaces is the fact that vectors can be
written in different orthonormal basis. In a vector space of $M$ dimensions
and an orthogonal vectors $\left\vert n\right\rangle $ we can construct $M$
orthogonal linear combinations as%
\begin{equation}
\left\vert \varphi _{n}\right\rangle
=\sum\limits_{j=1}^{M}c_{j}^{(n)}\left\vert j\right\rangle \text{ \ \ \ \ \ }%
j=1,2,...,M  \label{6}
\end{equation}%
where $c_{j}^{(n)}$ are the coefficients of the superposition and due to the
orthogonality $\left\langle \varphi _{m}\mid \varphi _{n}\right\rangle
=\delta _{nm}$ obey%
\begin{equation}
\sum\limits_{j=1}^{M}(c_{j}^{(m)})^{\ast }c_{j}^{(n)}=\delta _{nm}\text{ for 
}n\neq m  \label{7}
\end{equation}%
We can write Eq.(\ref{6}) as $\left\vert \varphi \right\rangle =U\left\vert
\varphi _{0}\right\rangle $, where $\left\vert \varphi \right\rangle =\left(
\left\vert \varphi _{1}\right\rangle \text{ }\left\vert \varphi
_{2}\right\rangle \text{ }\cdots \text{ }\left\vert \varphi
_{M}\right\rangle \right) ^{T}$ is the transformed basis as a row vector and 
$\left\vert \varphi _{0}\right\rangle =\left( \left\vert 1\right\rangle 
\text{ }\left\vert 2\right\rangle \text{ }\cdots \text{ }\left\vert
M\right\rangle \right) ^{T}$ is the original basis as a column vector and%
\begin{equation}
U=\left( 
\begin{array}{cccc}
c_{1}^{(1)} & c_{1}^{(2)} & ... & c_{M}^{(1)} \\ 
c_{2}^{(1)} & c_{2}^{(2)} & \cdots & c_{2}^{(M)} \\ 
\vdots & \vdots & \ddots & \vdots \\ 
c_{M}^{(1)} & c_{M}^{(2)} & \cdots & c_{M}^{(M)}%
\end{array}%
\right)  \label{7.1}
\end{equation}%
is a $M\times M$ matrix where each column contains the coefficients of the
linear combination $\left\vert \varphi ^{(n)}\right\rangle $. With this
notation for the transformation $U$, to satisfy the scalar product
invariance under transformation $U$ must obeys $U^{T}U=I$ where $U^{T}$ is
transpose of $U$. As it was shown in \cite{qcl}, the transformation $U$
belongs to the $SO(M)$ is the special orthogonal group in $M$ dimensions
with $M(M-1)/2$ generators of the Lie algebra \cite{geo}. For compact groups
such as $SO(M)$, the parameters of the Lie algebra are angles.\footnote{%
Symmetry considerations have been explored in econophysics, where the
different choice of basis of the vector space has been used to define
invariant matrix rates of returns \cite{anna}.} This parameter space
dimension is larger than the loan duration $M$ for $M>3$ which implies that
the transformation $U$ in the vector space of dimension $M$ provides a large
number of degrees of freedom to tune the payment schedule with better
benefits for the borrower. The transformation $U\in SO(M)$ induces a
transformation on any operator $O$ as $\overline{O}=UOU^{T}$. The loan
operators transform as 
\begin{eqnarray}
\overline{D} &=&UDU^{T}\text{ \ \ \ \ \ }\overline{A}=UAU^{T}
\label{7.1.1.1} \\
\overline{Y} &=&UYU^{T}\text{ \ \ \ \ }\overline{Q}=UQU^{T}  \notag
\end{eqnarray}%
The mean values of the transformed loan operators in the original basis read%
\begin{eqnarray}
\left\langle n\right\vert \overline{D}\left\vert n\right\rangle &=&\overline{%
d}_{n}\text{ \ \ \ \ \ \ \ }\left\langle n\right\vert \overline{Y}\left\vert
n\right\rangle =\overline{y}_{n}  \label{7.1.2} \\
\left\langle n\right\vert \overline{A}\left\vert n\right\rangle &=&\overline{%
a}_{n}\text{ \ \ \ \ \ \ }\left\langle n\right\vert \overline{Q}\left\vert
n\right\rangle =\overline{q}_{n}  \notag
\end{eqnarray}%
and give the expected values of the loan magnitudes at each period. Writing
the operators in the spectral decomposition, is not difficult to show that 
\begin{gather}
\overline{d}_{n}=\underset{j=1}{\overset{M}{\sum }}\left\vert
c_{j}^{(n)}\right\vert ^{2}d_{n}\text{ \ \ \ \ }\overline{a}_{n}=\underset{%
j=1}{\overset{M}{\sum }}\left\vert c_{j}^{(n)}\right\vert ^{2}a_{n}
\label{7.1.4} \\
\text{ }\overline{y}_{n}=\underset{j=1}{\overset{M}{\sum }}\left\vert
c_{j}^{(n)}\right\vert ^{2}y_{n}\text{ \ \ \ \ }\overline{q}_{n}=\underset{%
j=1}{\overset{M}{\sum }}\left\vert c_{j}^{(n)}\right\vert ^{2}q_{n}  \notag
\end{gather}%
From this point of view, the basis rotation mixes the mean values of the
classical loan and according to the time evolution of the loan, given by the
creation operator $a^{\dagger }$ (see Eq.(30) of Eq.(\cite{qcl})), we can
write $\overline{a}=UaU^{T}$ and it can be shown that $\overline{a}%
\left\vert \varphi _{n}\right\rangle =N_{n}\left\vert \varphi
_{n-1}\right\rangle $, where $\left\vert \varphi _{n-1}\right\rangle $ and $%
\left\vert \varphi _{n}\right\rangle $ are two orthogonal vectors obtained
from the original orthonormal basis by rotation, which means that $\overline{%
a}$ and $\overline{a}^{\dagger }$ acts as creation and annihilation
operators of loan configurations in the rotated basis. This implies that in
the rotated basis, the loan time evolution is 
\begin{equation}
\left\vert \varphi _{n}\right\rangle =\left(
\prod\limits_{j=2}^{n}N_{j}\right) ^{-1}(\overline{a}^{\dagger
})^{n-1}\left\vert \varphi _{1}\right\rangle  \label{7.1.6}
\end{equation}%
This last result is important because indicates how to obtain the
time-ordered rotated loan values. Eq.(\ref{7.1.4}) shows that the mean
values of the loan operators in the new orthogonal basis do not obey the
recurrence relations of Eq.(\ref{c1}), that is, $\overline{d}_{n}\neq t%
\overline{y}_{n-1}$, $\overline{d}_{n+1}\neq \overline{d}_{n}-\overline{a}%
_{n+1}$ but $\overline{q}_{n}=\overline{y}_{n}+\overline{a}_{n}$. This is
expected because it is the algebra given in Eq.(\ref{new1}) what truly
represents the loan structure and not any particular representation of the
loan operators in an orthonormal basis.

The indexed loans can be written in terms of the GHA by recalling Eq.(\ref%
{c2}), where the interest rate $t_{n}$ depends on $n$. In this case the
interest rate operator $T$ is not degenerated. There are particular indexed
loans that can be obtained from constant interest rate by creating a debt in
an non-monetary unit (see for example Sect. 4 of \cite{qcl}). For instance,
by defining $t_{n}=(1+t)\frac{\alpha _{n}}{\alpha _{n-1}}-1$, where $\alpha
_{n}$ is some arbitrary function of $n$ and using Eq.(\ref{c2}), we obtain $%
\overline{d}_{n}=(1+t)\overline{d}_{n-1}-\overline{q}_{n}$ with $\overline{q}%
_{n}=\alpha _{n}q$ and $\overline{d}_{n}=\frac{d_{n}}{\alpha _{n}}$. The new
debt in monetary units $\overline{d}_{n}$ is proportional to the debt $d_{n}$
in non-monetary units. The variable $\alpha _{n}$ can be related to
macroeconomical variables such as inflation. In \cite{qcl} we have shown
that the rotation of the orthogonal basis provides us with a solution to the
loan payments increment by choosing the specific angles so that the payments
remain constant.\ These results are useful when inflation volatility has an
effect on the allocation of bank loans where bank managers behave more
conservatively \cite{mustafa}. Implementing bank loans on vectorial spaces
reduces the inflation risk due to the volatility and in turn the borrower
evades the default. In general in the case of borrower default, the debt
must be refinanced and the terms and conditions vary for different countries
and banking regulations. Under financial distress, a debt obligation can be
replaced by another debt, which implies debt restructuring, and where it is
necessary to reduce the inherent risk or to reduce the monthly repayment
amount. These refinancing has a penalty that implies the borrower will have
to take longer to pay off the debt, altering the maturity. Writing debt
obligations in terms of an algebra of operators gives the ability to
manipulate the orthogonal directions of the vectorial space and to
reconfigure the payments without altering the interest rate or the loan
duration, and this is clearly an improvement to this financial instrument.

\subsubsection{Superposition of classical amortization systems}

Let us consider the rotated amortization and periodic payment values $%
\overline{a}_{n}=\underset{j=1}{\overset{M}{\sum }}\left\vert
c_{j}^{(n)}\right\vert ^{2}a_{j}$ and $\overline{q}_{n}=\underset{j=1}{%
\overset{M}{\sum }}\left\vert c_{j}^{(n)}\right\vert ^{2}q_{j}$. As it was
shown in the last section, in the French amortization system, the periodic
payment is constant $q_{n}=q_{F}$. Then, the rotated values read $\overline{q%
}_{n}=q_{F}$, that is, although the basis is rotated, a diagonal matrix with
identical eigenvalues is invariant under rotations as it is occur with the $%
Q $ payment matrix in the French system. This implies that we can change the
payment schedule in those amortization systems with non-constant periodic
payments. From last equation we can see that $\overline{q}_{n}$ can be
constant when we choose $c_{j}^{(n)}=1/\sqrt{M}$ and the rotated periodic
payments are $\overline{q}_{n}=\overline{q}=\frac{1}{M}\underset{j=1}{%
\overset{M}{\sum }}q_{j}$, which is the mean value of the original
non-constant periodic payments. Then, an amortization system with
non-constant periodic payments can be transformed into a constant periodic
payment amortization system with a specific rotation matrix $U$. For
example, the German system with constant amortization and decreasing
periodic payments $q_{n}^{(G)}=\frac{d_{0}}{M}+t\frac{d_{0}}{M}(M-n+1)$ can
be turned to a French amortization system with a constant periodic payment%
\begin{equation}
\overline{q}=\frac{1}{M}\underset{j=1}{\overset{M}{\sum }}q_{j}^{(G)}=\frac{%
d_{0}}{2M}\left( t+Mt+2\right)  \label{su2}
\end{equation}%
Nevertheless, the last result is not identical to the French periodic
payment $q_{F}$ indicating that the rotation of a German system can give a
German$-$French system with constant amortization and periodic payment, but
this superposition is different from the one obtained by rotating the French
system. This new mixed French-German system loan cannot be obtained by the
usual formalism of Eq.(\ref{c1}) because if $q_{n}$ and $a_{n}$ are
constants then $y_{n}$ and $d_{n}$ cannot change. In this sense, the
relation between writing the amortization systems on vector spaces and the
usual description is analogue to the relation between quantum game theory
and classical game theory, where in the quantum theory, the set of possible
strategies is enlarged by allowing superposition of strategies (\cite{pio0}
and \cite{flit}). The same procedure is obtained with the introduction of
the vector space, which provides a large degrees of freedom encoded in the
distance-preserving rotation.

\subsection{Entangled loans}

\begin{figure}[h]
\centering\includegraphics[width=50mm,height=20mm]{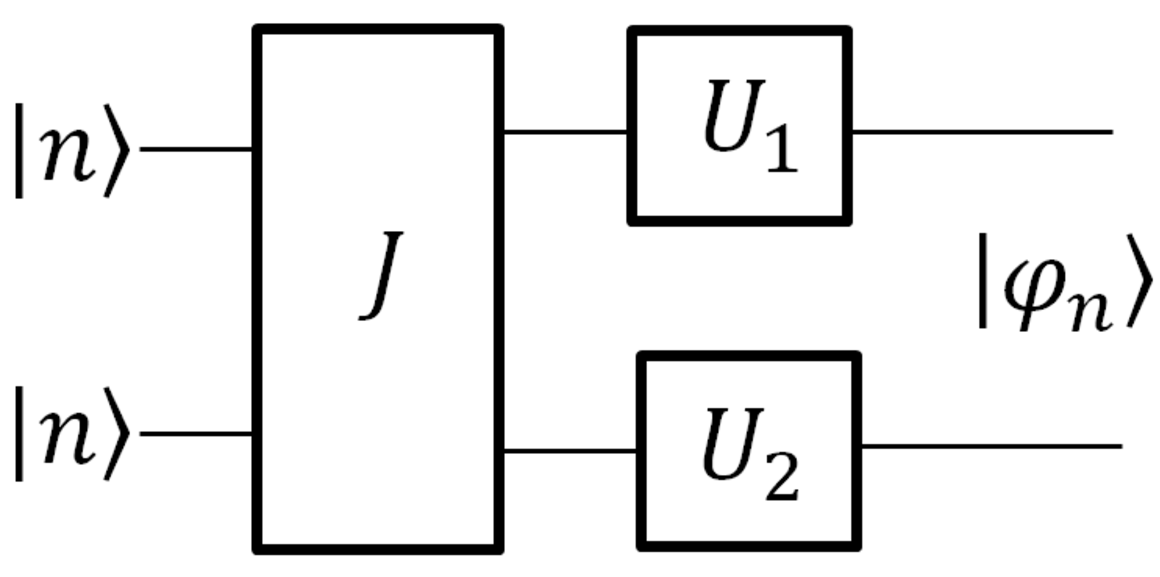}
\caption{The setup of two-borrower entanglement loan.}
\label{statetr}
\end{figure}

\begin{figure}[h]
\centering\includegraphics[width=150mm,height=50mm]{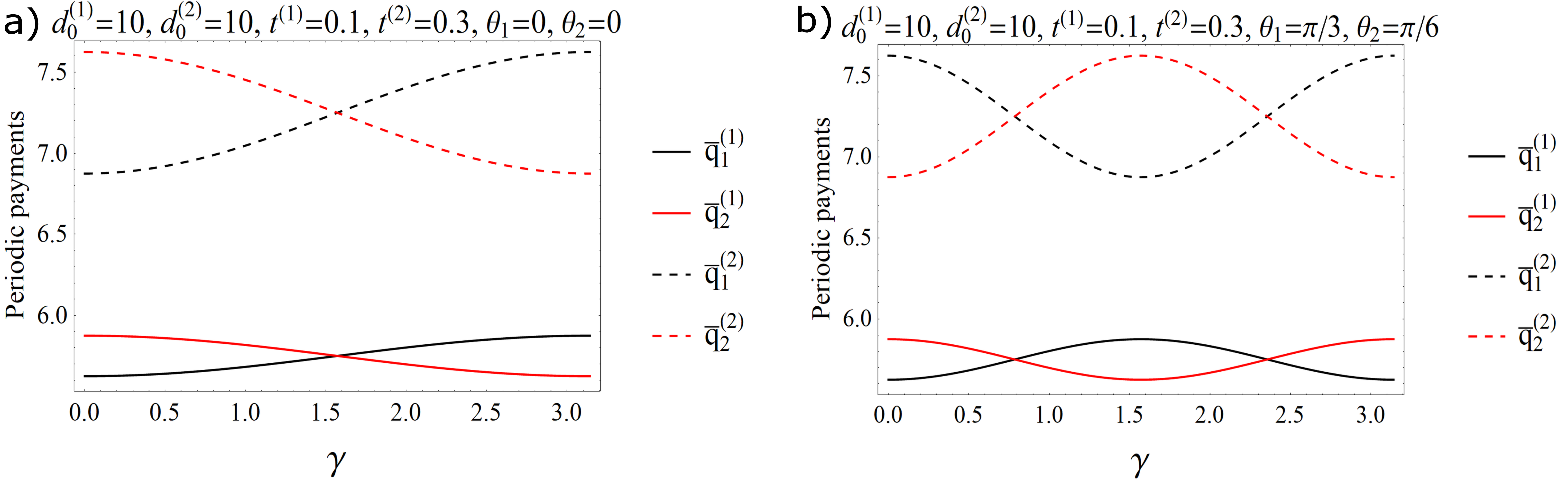}
\caption{Entangled periodic payments as a function of $\protect\gamma $ for $%
M=2$. Both borrowers have identical initial debts but different interest
rate. a) $\protect\theta _{1}=0$ and $\protect\theta _{2}=0$. b)\ $\protect%
\theta _{1}=\protect\pi /3$ and $\protect\theta _{2}=\protect\pi /6$.}
\label{entan2}
\end{figure}

\begin{figure}[h]
\centering\includegraphics[width=80mm,height=20mm]{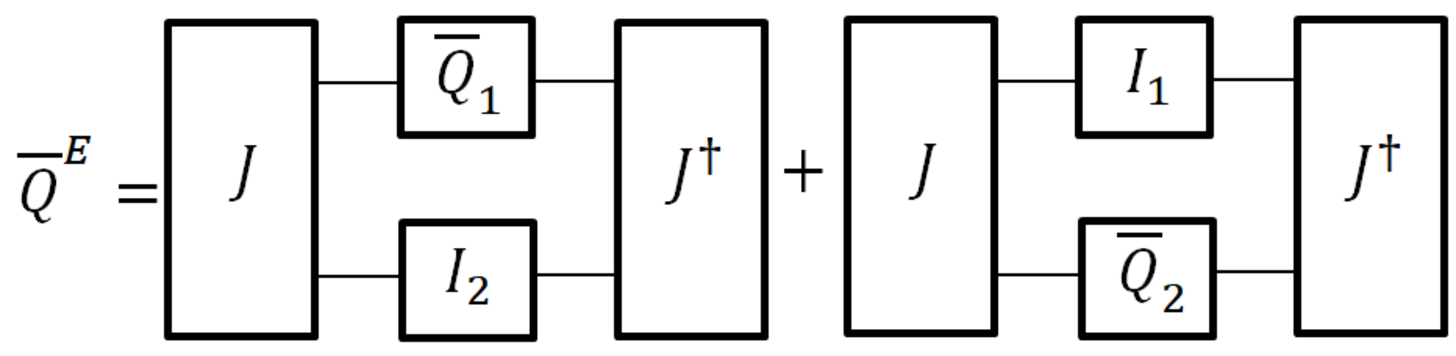}
\caption{The setup of two-borrower entanglement loan.}
\label{Qenta}
\end{figure}

\begin{figure}[h]
\centering\includegraphics[width=150mm,height=94mm]{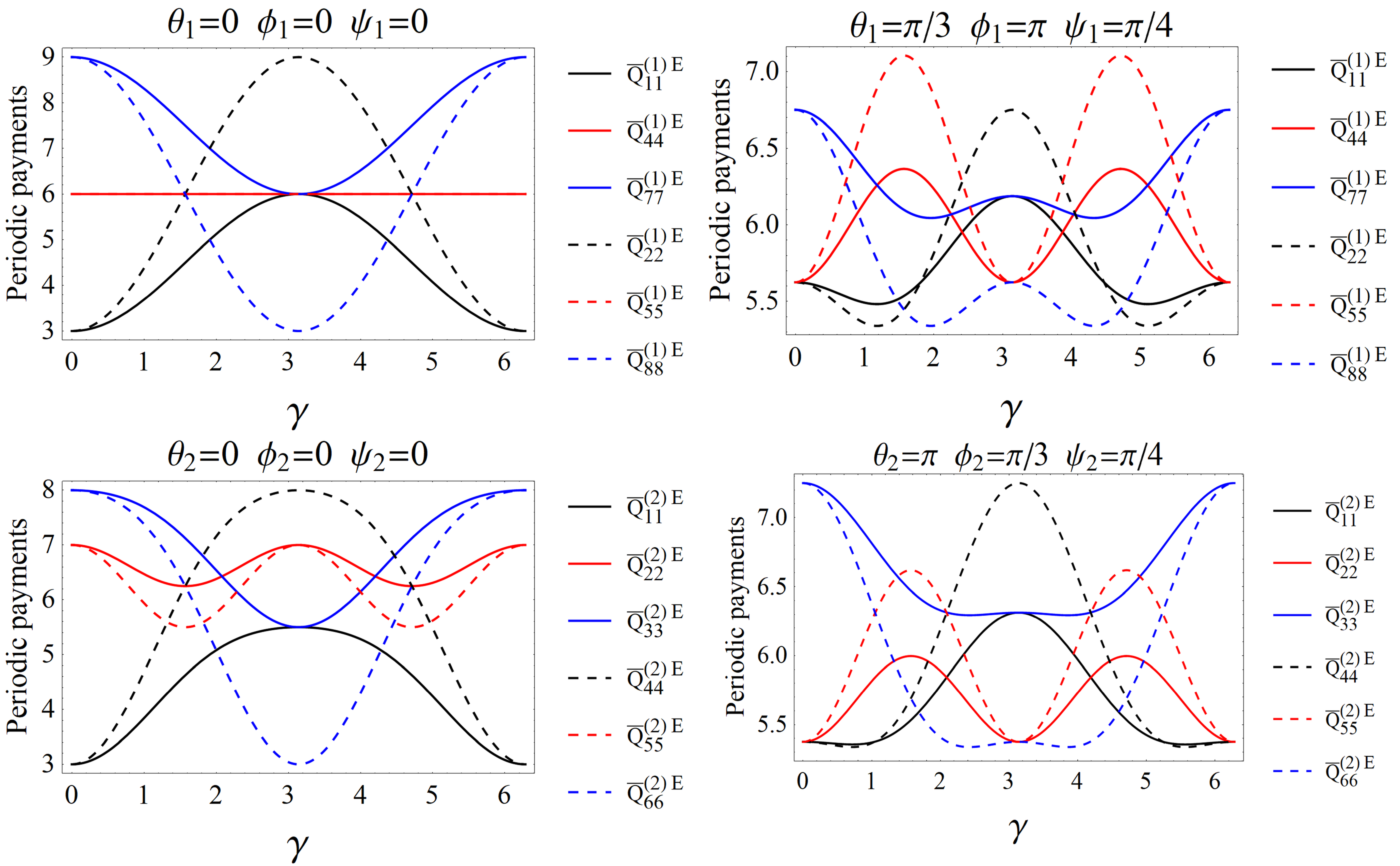}
\caption{Entangled payments for the first borrower as a function of $\protect%
\gamma $ for $M=3$ and two specific set of angles. The initial payments are $%
q_{1}^{(1)}=3$, $q_{2}^{(1)}=6$ and $q_{3}^{(1)}=9$. a) $\protect\theta %
_{1}=0$, $\protect\phi _{1}=0$, $\protect\psi _{1}=0$. b) $\protect\theta %
_{1}=\protect\pi /3$, $\protect\phi _{1}=\protect\pi $, $\protect\psi _{1}=%
\protect\pi /4$. c) $\protect\theta _{2}=0$, $\protect\phi _{2}=0$, $\protect%
\psi _{2}=0$. d) $\protect\theta _{2}=\protect\pi /3$, $\protect\phi _{2}=%
\protect\pi $, $\protect\psi _{2}=\protect\pi /4$.}
\label{entan31}
\end{figure}

To explore the consequences of rewriting the amortization systems on
vectorial spaces, we can analyze how we can combine two loans using the
tensor product of the vector spaces. For simplicity we will analyze the most
elemental entanglement loans with $M=2$ and initial debts $d_{0}^{(i)}$ and
interest $t^{(i)}$ with $i=1,2$ denoting the loans. Later we will generalize
the results to arbitrary $M$. The vector space of the combined loans is $%
\mathcal{H}_{1}\otimes \mathcal{H}_{2}$. An orthonormal basis of this
vectorial space can be written as $\{\left\vert s_{1},s_{2}\right\rangle
=\left\vert s_{1}\right\rangle \otimes \left\vert s_{2}\right\rangle \}$
with $s_{1},s_{2}=1,2$. The total payment matrix can be written as 
\begin{equation}
Q=Q_{1}\otimes I_{2}+I_{1}\otimes Q_{2}  \label{enta1}
\end{equation}%
where $I_{1}$ and $I_{2}$ are the identity operators in $\mathcal{H}_{1}$and 
$\mathcal{H}_{2}$ and $Q_{1}$ and $Q_{2}$ are the payment matrices in each
vectorial space. We can compute the diagonal elements of the payment matrix
of each loan as%
\begin{equation}
q_{n}^{(1)}=\left\langle n,n\right\vert Q_{1}\otimes I_{2}\left\vert
n,n\right\rangle \text{ \ \ \ \ \ }q_{n}^{(2)}=\left\langle n,n\right\vert
I_{1}\otimes Q_{2}\left\vert n,n\right\rangle   \label{enta2}
\end{equation}%
where $\left\vert n,n\right\rangle $ is the configuration state of the
combined loan. The total diagonal elements of the payment matrix can be
computed as 
\begin{equation}
q_{n}=\left\langle n,n\right\vert Q_{1}\otimes I_{2}+I_{1}\otimes
Q_{2}\left\vert n,n\right\rangle =q_{n}^{(1)}+q_{n}^{(2)}  \label{enta3}
\end{equation}%
To generate loan entanglement configurations we must first consider that
each borrower is allowed to apply an unitary transformation over its own
vectorial space and we can write $U=U_{1}\otimes U_{2}$ where $U_{1/2}$
belongs to the $SO(2)$ symmetry group of rotations acting on $\mathcal{H}%
_{1/2}$ respectively. But if this transformation is done over the
configuration state $\left\vert n,n\right\rangle $ we obtain that each loan
behaves independently of the other. To avoid this triviality, we can recall
the strategy of the quantum prisoner dilemma, where the initial state is a
shared qubit between the prisoners and a unitary operator $J$ entangles the
two configuration states before the application of strategies (\cite{eisert}
and \cite{marek}). Following the same procedure, a suitable entanglement
operator can be chosen for $M=2$ as 
\begin{equation}
J_{2}=e^{i\gamma \sigma _{x}^{(2)}\otimes \sigma _{x}^{(2)}}=\cos (\gamma
)I_{1}^{(2)}\otimes I_{2}^{(2)}+i\sin (\gamma )\sigma _{x}^{(2)}\otimes
\sigma _{x}^{(2)}  \label{enta3.1}
\end{equation}%
where $\sigma _{x}^{(2)}=\left( 
\begin{array}{cc}
0 & 1 \\ 
1 & 0%
\end{array}%
\right) $ and $\gamma $ is the a measure of the loan entanglement. The
unitary operator $J_{2}$ is known for both borrowers and is symmetric with
respect to the interchange of the two borrowers. The transformed loan
configuration reads%
\begin{equation}
\left\vert \psi _{n}\right\rangle =U_{1}\otimes U_{2}J\left\vert
n,n\right\rangle   \label{enta4}
\end{equation}%
with $n=1,2$. In Fig. \ref{statetr}, a physical model of the transformed
loan configuration of last equation is shown where each borrower has a qubit
and can manipulate it independently after a gate application $J$ that
produce an entangled state. 
The entangled diagonal elements of the transformed payment matrix can be
written as%
\begin{equation}
\overline{q}_{n}^{(1)E}=\left\langle \psi _{n}\right\vert Q_{1}\otimes
I_{2}\left\vert \psi _{n}\right\rangle \text{ \ \ \ \ \ }\overline{q}%
_{n}^{(2)E}=\left\langle \psi _{n}\right\vert I_{1}\otimes Q_{2}\left\vert
\psi _{n}\right\rangle   \label{enta5}
\end{equation}%
By using Eqs.(\ref{enta3.1}) and (\ref{enta4}), last equation can be written
for $M=2$ as 
\begin{equation}
\overline{q}_{n}^{(j)E}=\cos ^{2}(\gamma )\overline{q}_{n}^{(j)}+\sin
^{2}(\gamma )\left\langle n\right\vert \sigma
_{x}^{(j)}U_{j}^{T}Q_{j}U_{j}\sigma _{x}^{(j)}\left\vert n\right\rangle 
\label{enta5.1}
\end{equation}%
where $\overline{q}_{n}^{(j)}=\left\langle n\right\vert U_{j}^{\dagger
}Q_{j}U_{j}\left\vert n\right\rangle $ is the transformed periodic payment
of the $j$ borrower. Introducing identity operators $I$ between $\sigma
_{x}^{(j)}U_{j}^{T}$ and $U_{j}\sigma _{x}^{(j)}$, $\overline{q}_{n}^{(j)E}$
can be written in compact form as%
\begin{equation}
\overline{q}_{n}^{(j)E}=\cos ^{2}(\gamma )\overline{q}_{n}^{(j)}+\sin
^{2}(\gamma )\left\langle n\right\vert \sigma _{x}^{(j)}\overline{Q}%
_{j}\sigma _{x}^{(j)}\left\vert n\right\rangle   \label{enta5.2}
\end{equation}%
where $\overline{Q}_{j}=U_{j}^{T}Q_{j}U_{j}$. Using $n=1,2$ and Eq.(36) of 
\cite{qcl} is not difficult to show that for $M=2$ is%
\begin{eqnarray}
\overline{q}_{1}^{(i)E} &=&\frac{1}{2}%
(q_{1}^{(i)}+q_{2}^{(i)}+(q_{1}^{(i)}-q_{2}^{(i)})\cos \gamma \cos \theta
_{i})  \label{enta6} \\
\overline{q}_{2}^{(i)E} &=&\frac{1}{2}%
(q_{1}^{(i)}+q_{2}^{(i)}-(q_{1}^{(i)}-q_{2}^{(i)})\cos \gamma \cos \theta
_{i})  \notag
\end{eqnarray}%
where $i=1,2$. In Fig. (\ref{entan2}) the two periodic payments are shown as
a function of $\gamma $ for different values of $\theta $. Is not difficult
to show that\ $\overline{q}_{1}^{(i)E}+\overline{q}%
_{2}^{(i)E}=q_{1}^{(i)}+q_{2}^{(i)}$ as it is expected. In this figure, both
borrowers must repay an identical initial debts but with different interest.
In figure (\ref{entan2})a), both borrowers do not rotate their vectorial
spaces but the entanglement parameter can be chosen in such a way to obtain
increasing payments in time for both borrowers ($\gamma <\pi /2$) or
decreasing payments ($\gamma >\pi /2$) or constant payments ($\gamma =\pi /2$%
). Interestingly is when both borrowers selects $\gamma $ but the first
rotate $\theta _{1}=\pi /3$ and simultaneously the second rotate $\theta
_{2}=\pi /6$ at the beginning of the loan (see Fig. (\ref{entan2})b)). In
this case, when one payment schedule is decreasing the other is increasing
and vice versa. Although this example is simple, it shows how to obtain
different payment schedules for both borrowers in which each of them can
choose its own strategy of repayment. 
To generalize the $M=2$ case to arbitrary $M$, where each loan has identical
maturity, we can write the entanglement operator as%
\begin{equation}
J_{M}=e^{i\gamma \sigma _{x}^{(M)}\otimes \sigma _{x}^{(M)}}  \label{enta7}
\end{equation}%
where $\sigma _{x}^{(M)}$ is the $x$ component of the angular momentum in
the $M\times M$ representation chosen where $\sigma _{z}$ is diagonal. That
is, we can consider a vectorial space of dimension $M$ and we can consider
the $2j+1=M$ representation of the angular momentum $j$ over the vectorial
space. The loan configurations $\left\vert n\right\rangle $ from $n=1$ to $M$
are collinear to the $j=(M-1)/2$ spin projection states. For instance, with $%
j=1/2$, the spin projection states to the $z$ axis are $\left\vert j=\frac{1%
}{2},m=-\frac{1}{2}\right\rangle $ and $\left\vert j=\frac{1}{2},m=\frac{1}{2%
}\right\rangle $ and correspond to the loan configurations $\left\vert
1\right\rangle $ and $\left\vert 2\right\rangle $ with $M=2$. In a similar
way, for $j=1$, the spin projection states $\left\vert j=1,m=1\right\rangle $%
, $\left\vert j=1,m=0\right\rangle $ and $\left\vert j=1,m=-1\right\rangle $
correspond to the loan states $\left\vert 1\right\rangle $, $\left\vert
2\right\rangle $ and $\left\vert 3\right\rangle $ respectively for $M=3$.
The entanglement operator $\sigma _{x}^{(M)}\otimes \sigma _{x}^{(M)}$ is an
arbitrary choice since we can also choose $i\sigma _{y}^{(M)}\otimes i\sigma
_{y}^{(M)}$ or any other qudit entanglement gate such as the SWAP or CNOT\
gate \cite{alti}. The entangled payment matrix for each borrower can be
written as%
\begin{eqnarray}
\overline{Q}^{(1)E} &=&J_{M}^{\dagger }(U_{1}^{T}\otimes
U_{2}^{T})(Q_{1}\otimes I_{2})(U_{1}\otimes U_{2})J_{M}  \label{enta7.1} \\
\overline{Q}^{(2)E} &=&J_{M}^{\dagger }(U_{1}^{T}\otimes
U_{2}^{T})(I_{1}\otimes Q_{2})(U_{1}\otimes U_{2})J_{M}  \notag
\end{eqnarray}%
where $J^{\dagger }=(J^{T})^{\ast }$ is the adjoint of $J_{M}$. From last
equation it can be seen that $\overline{Q}^{(1)E}=J_{M}^{\dagger }(\overline{%
Q}_{1}\otimes I_{2})J_{M}$ and $\overline{Q}^{(2)E}=J_{M}^{\dagger
}(I_{1}\otimes \overline{Q}_{2})J_{M}$. In Fig. \ref{Qenta}, a setup for the
entangled payment matrix is shown where each borrower can rotate its own
payment matrix between the entangling and disentangling operators $J$ and $%
J^{\dagger }$. 
The matrix $J_{M}$ can be written using spin matrix polynomials (see \cite%
{curt}) and by the Cayley$-$Hamilton theorem, the expansion is a finite sum
of powers of $\sigma _{x}^{(M)}\otimes \sigma _{x}^{(M)}$ where the highest
power is of order $M-1$ 
\begin{equation}
e^{i\gamma \sigma _{x}^{(M)}\otimes \sigma
_{x}^{(M)}}=\sum\limits_{k=0}^{M-1}\frac{c_{k}(\gamma )}{k!}(2i\sin (\frac{%
\gamma }{2}))^{k}(\sigma _{x}^{(M)})^{k}\otimes (\sigma _{x}^{(M)})^{k}
\label{enta7.1.1}
\end{equation}%
where $c_{k}(\gamma )$ is a some function of $\gamma $ (see Eq.(2) of \cite%
{curt}). For construction $Tr(\overline{Q}^{(1)E})=Tr[\overline{Q}%
_{1}\otimes I_{2}]=Tr[Q_{1}]Tr[I_{2}]=MTr[Q_{1}]$ and similar for $Tr(%
\overline{Q}^{(2)E})=MTr[Q_{2}]$. This indicates that there will be
different configurations of entangled payments for each borrower when $M>2$.
To be precise about the different configurations, let us consider $M=3$,
then the $j=1$ matrix representation of the $x$ component of the angular
momentum is 
\begin{equation}
\sigma _{x}^{(3)}=\frac{1}{\sqrt{2}}\left( 
\begin{array}{ccc}
0 & 1 & 0 \\ 
1 & 0 & 1 \\ 
0 & 1 & 0%
\end{array}%
\right)   \label{enta7.2}
\end{equation}%
then the entanglement matrix $J_{3}$ reads%
\begin{equation}
J_{3}=I_{1}\otimes I_{2}+i\sin (\gamma )\sigma _{x}^{(3)}\otimes \sigma
_{x}^{(3)}+[\cos (\gamma )-1](\sigma _{x}^{(3)})^{2}\otimes (\sigma
_{x}^{(3)})^{2}  \label{enta7.3}
\end{equation}%
where we have used that $(\sigma _{x}^{(3)})^{3}=\sigma _{x}^{(3)}$.
Computing $\overline{Q}^{(1)E}$ and $\overline{Q}^{(2)E}$ using Eq.(\ref%
{enta7.1}), it can be shown that the diagonal matrix elements obey%
\begin{eqnarray}
\overline{Q}_{11}^{(1)E} &=&\overline{Q}_{33}^{(1)E}\text{ \ \ \ \ }%
\overline{Q}_{44}^{(1)E}=\overline{Q}_{66}^{(1)E}\text{ \ \ \ \ \ }\overline{%
Q}_{77}^{(1)E}=\overline{Q}_{99}^{(1)E}  \label{enta7.4} \\
\overline{Q}_{11}^{(2)E} &=&\overline{Q}_{77}^{(2)E}\text{ \ \ \ \ }%
\overline{Q}_{22}^{(2)E}=\overline{Q}_{88}^{(2)E}\text{ \ \ \ \ \ }\overline{%
Q}_{33}^{(2)E}=\overline{Q}_{99}^{(2)E}  \notag
\end{eqnarray}%
and 
\begin{equation}
Tr[Q_{1}]=\overline{Q}_{11}^{(1)E}+\overline{Q}_{44}^{(1)E}+\overline{Q}%
_{77}^{(1)E}=\overline{Q}_{22}^{(1)E}+\overline{Q}_{55}^{(1)E}+\overline{Q}%
_{88}^{(1)E}  \label{enta7.5}
\end{equation}%
which are a consequence of $Tr[\overline{Q}_{i}^{E}]=3Tr[Q_{1}]$.
Identically for the second borrower we obtain%
\begin{equation}
Tr[Q_{2}]=\overline{Q}_{11}^{(2)E}+\overline{Q}_{22}^{(2)E}+\overline{Q}%
_{33}^{(2)E}=\overline{Q}_{44}^{(2)E}+\overline{Q}_{55}^{(2)E}+\overline{Q}%
_{66}^{(2)E}  \label{enta7.6}
\end{equation}%
where $\overline{Q}_{nn}^{(i)E}=\left\langle n,n\right\vert \overline{Q}%
_{i}^{E}\left\vert n,n\right\rangle $ is the mean value of $\overline{Q}%
_{i}^{E}$ in the loan configuration $\left\vert n,n\right\rangle \,$.Then we
have two entangled configurations payments for each borrower and is not
difficult to show that the number of entangled configurations is $M-1$ for
arbitrary $M$. The first borrower can choose the payments $\{\overline{Q}%
_{11}^{(1)E},\overline{Q}_{44}^{(1)E},\overline{Q}_{77}^{(1)E}\}$ or $\{%
\overline{Q}_{22}^{(1)E},\overline{Q}_{55}^{(1)E},\overline{Q}_{88}^{(1)E}\}$
and the second borrower can choose the payments $\{\overline{Q}_{11}^{(2)E},%
\overline{Q}_{22}^{(2)E},\overline{Q}_{33}^{(2)E}\}$ or $\{Q_{44}^{(2)E},%
\overline{Q}_{55}^{(2)E},\overline{Q}_{66}^{(2)E}\}$. In Fig. \ref{entan31},
the payments are shown as a function of the entanglement parameter $\gamma $
for different values of $\theta $, $\phi $ and $\psi $, where we have used
that $q_{1}^{(1)}=3$, $q_{2}^{(1)}=6$ and $q_{3}^{(1)}=9$ for the first loan
and $q_{1}^{(2)}=3$, $q_{2}^{(2)}=7$ and $q_{3}^{(2)}=8$. These values can
be obtained from a suitable choice of $t^{(i)}$ and $d_{0}^{(i)}$. In this
figure the thick lines are the first set and the dashed lines correspond to
the second set of payment for each borrower and the colour indicates payment
order (black line is the first payment, red line is the second payment and
blue line indicates third payment). 
The entangled payments are strongly correlated and once $\gamma $ is chosen,
each borrower can apply its own rotation altering the payment schedule.
Interestingly we can explore two loans with different maturities. For
simplicity $M_{1}=2$ and $M_{2}=3$ and the entanglement matrix can be
written as%
\begin{equation}
J=e^{i\gamma \sigma _{x}^{(2)}\otimes \sigma _{x}^{(3)}}=I_{1}\otimes
I_{2}+[\cos (\gamma )-1]I_{1}\otimes (\sigma _{x}^{(3)})^{2}+i\sin (\gamma
)\sigma _{x}^{(2)}\otimes \sigma _{x}^{(3)}  \label{enta7.7}
\end{equation}%
In this case, we obtain two configurations of payment schedules for $M=3$
and one configuration for $M=2$. For example, with $\phi _{1}=0$ and $\theta
_{2}=\phi _{2}=\psi _{2}=0$ the plots are given by Figs. \ref{entan2} and %
\ref{entan31}.

Should be stressed that a different description can be done for entangled
loans using density operators $\rho _{s}=\left\vert s,s\right\rangle
\left\langle s,s\right\vert $ with $s=1,2$ and the periodic payments of each
borrower in this configuration state $\rho _{s}$ can be computed with the
partial traces of $\rho _{s}$, which give the reduced density operator on
each vectorial space%
\begin{equation}
\rho _{s}^{(1)}=Tr_{2}\rho =\sum\limits_{s_{2}=1}^{2}\left\langle
s_{2}\right\vert \rho \left\vert s_{2}\right\rangle =\left\vert
s\right\rangle \left\langle s\right\vert \text{ \ \ \ \ \ \ \ }\rho
_{s^{\prime }}^{(2)}=Tr_{1}\rho =\sum\limits_{s_{1}=1}^{2}\left\langle
s_{1}\right\vert \rho \left\vert s_{1}\right\rangle =\left\vert s^{\prime
}\right\rangle \left\langle s^{\prime }\right\vert  \label{enta8}
\end{equation}%
where $\rho _{s}^{(1)}=$ $\left\vert s\right\rangle \left\langle
s\right\vert $ acts on $\mathcal{H}_{1}$ and $\rho _{s^{\prime }}^{(2)}=$ $%
\left\vert s^{\prime }\right\rangle \left\langle s^{\prime }\right\vert $
acts on $\mathcal{H}_{2}$. The periodic payments can be computed as%
\begin{equation}
q_{s}^{(1)}=Tr(\rho _{s}^{(1)}Q_{1})\text{ \ \ \ \ \ \ \ \ }%
q_{s}^{(2)}=Tr(\rho _{s}^{(2)}Q_{2})  \label{enta9}
\end{equation}%
where $Q_{j}=\sum\limits_{s=1}^{2}q_{s}^{(j)}\left\vert s_{j}\right\rangle
\left\langle s_{j}\right\vert $ is the periodic payment operator of the $j$
loan. This description is suitable to define an entanglement measure between
the loans. There are several operational entanglement measures, for example
distillable entanglement, distillable key and entanglement cost, as well as
abstractly defined measures such as concurrence or negativity \cite{measu}.
When the overall loan state is pure, the entanglement entropy is a suitable
measure defined as $S=-Tr(\rho \ln \rho )$ where $\rho $ can be any reduced
state or the total density operator. For instance, considering $M=2$ we have
two loan configurations%
\begin{eqnarray}
\left\vert \varphi _{1}\right\rangle &=&\cos (\gamma )U_{1}\left\vert
1\right\rangle \otimes U_{2}\left\vert 1\right\rangle +i\sin (\gamma
)U_{1}\left\vert 2\right\rangle \otimes U_{2}\left\vert 2\right\rangle
\label{enta10} \\
\left\vert \varphi _{2}\right\rangle &=&\cos (\gamma )U_{1}\left\vert
2\right\rangle \otimes U_{2}\left\vert 2\right\rangle +i\sin (\gamma
)U_{1}\left\vert 1\right\rangle \otimes U_{2}\left\vert 1\right\rangle 
\notag
\end{eqnarray}%
by writing $\rho _{1}=\left\vert \varphi _{1}\right\rangle \left\langle
\varphi _{1}\right\vert $ and $\rho _{2}=\left\vert \varphi
_{2}\right\rangle \left\langle \varphi _{2}\right\vert $ and computing the
partial trace over the second borrower we obtain%
\begin{gather}
\widetilde{\rho }_{1}^{(1)}=Tr_{2}\rho _{1}=\cos ^{2}\gamma U_{1}\left\vert
1\right\rangle \left\langle 1\right\vert U_{1}^{T}+\sin ^{2}\gamma
U_{1}\left\vert 2\right\rangle \left\langle 2\right\vert U_{1}^{T}=
\label{enta11} \\
\left( 
\begin{array}{cc}
\cos ^{2}\gamma \cos ^{2}\theta _{1}+\sin ^{2}\gamma \sin ^{2}\theta _{1} & 
(\cos ^{2}\gamma -\sin ^{2}\gamma )\sin \theta _{1}\cos \theta _{1} \\ 
(\cos ^{2}\gamma -\sin ^{2}\gamma )\sin \theta _{1}\cos \theta _{1} & \cos
^{2}\gamma \sin ^{2}\theta _{1}+\sin ^{2}\gamma \cos ^{2}\theta _{1}%
\end{array}%
\right)  \notag
\end{gather}%
and 
\begin{gather}
\widetilde{\rho }_{2}^{(1)}=Tr_{2}\rho _{2}=\cos ^{2}\gamma U_{1}\left\vert
2\right\rangle \left\langle 2\right\vert U_{1}^{T}+\sin ^{2}\gamma
U_{1}\left\vert 1\right\rangle \left\langle 1\right\vert U_{1}^{T}=
\label{enta12} \\
\left( 
\begin{array}{cc}
\cos ^{2}\gamma \sin ^{2}\theta _{1}+\sin ^{2}\gamma \cos ^{2}\theta _{1} & 
(\sin ^{2}\gamma -\cos ^{2}\gamma )\cos \theta _{1}\sin \theta _{1} \\ 
(\sin ^{2}\gamma -\cos ^{2}\gamma )\cos \theta _{1}\sin \theta _{1} & \cos
^{2}\gamma \cos ^{2}\theta _{1}+\sin ^{2}\gamma \sin ^{2}\theta _{1}%
\end{array}%
\right)  \notag
\end{gather}%
where the subindex indicates the payment and the superindex indicates the
borrower. 
\begin{figure}[h]
\centering\includegraphics[width=100mm,height=60mm]{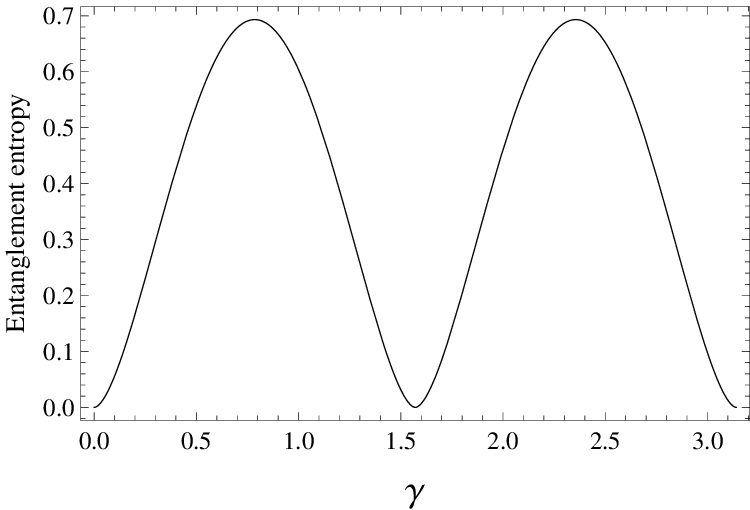}
\caption{Entanglement entropy as a function of the entanglement parameter $%
\protect\gamma $ for the reduced states of two $M=2$ loans. }
\label{entropy}
\end{figure}
Both reduced states for the first borrower $\widetilde{\rho }_{1}^{(1)}$ and 
$\widetilde{\rho }_{2}^{(1)}$ gives the same entanglement entropy and due to
the symmetry with respect to the interchange of the two borrowers the same
result is obtained for the second borrower%
\begin{equation}
S=S(\widetilde{\rho }_{i}^{(j)})=-\cos ^{2}\gamma \ln (\cos ^{2}\gamma )-\ln
(\sin ^{2}\gamma )\sin ^{2}\gamma  \label{enta13}
\end{equation}%
which is identical to the result obtained in \cite{duxu}. This result can be
obtained using the bi-orthogonal decomposition of $\widetilde{\rho }%
_{j}^{(l)}$ , where the diagonal coefficients are $\cos ^{2}\gamma $ and $%
\sin ^{2}\gamma $ respectively. In Fig. \ref{entropy}, $S$ is shown as a
function of $\gamma $, where the upper bound is obtained for $\gamma =\frac{%
\pi }{4}$ and $\gamma =\frac{3}{4}\pi $ and no entanglement for $\gamma =\pi
/2$, where $J=i\sigma _{x}\otimes \sigma _{x}$. As it is expected, $S$ does
not depend on the rotation angles and the strategies of the borrowers do not
change the entanglement measure. In turn, it can be shown that the
entanglement entropy of $\rho _{i}$ gives $S(\rho _{i})=0$ because $\rho
_{i} $ is a pure state. This implies that the mutual information defined as 
\begin{equation}
I=S(\rho _{i}^{(1)})+S(\rho _{i}^{(2)})-S(\rho _{i})=2S  \label{enta14}
\end{equation}%
which express the fact that the information is stored in the reduced states
and not in the composite loan configuration, which is maximally determined.

\subsection{Final discussion}

The linear algebra applied to the amortization system can be used to explore
how to redesign mortgage loans with a long duration such as $M=360$ (30
years loan) or less. This large maturity implies that the borrower is
exposed to macrovolatility. Different designs of countercyclical payments
have been studied where fixed-rate mortgages can be converted to
adjustable-rate mortgages \cite{adam}. These designs can be redefined in
terms of the $SO(M)$ symmetry that introduces $M(M-1)/2$ angles. A large
classification of subgroups of $SO(M)$ can be found in the literature in
terms of cosets and conjugacy classes (see \cite{geo} chapter 19 to 25) and
these subgroups can be used to define rotations that act only on a specific
number of payments. This can be achieved by selecting the specific
generators of the rotations and by applying the exponential map. For large $%
M $, this implies that the real application of loans with amortization
systems defined on vector spaces includes the development of software to
model the rotations applied by the borrower throughout the repayment. The
large parameter space of the Lie algebra involves the manipulation of $\
M(M-1)/2$ rotation angles at once and this can be computationally expensive.
The algebra of operators defined in Eq.(\ref{new1}) is the most simple one
to obtain the recurrence relations of the amortization system and is very
restrictive with respect the commutativity of the loan operators, which must
be compatible. This could be relaxed via non-commutative operators, but it
would imply the impossibility of a joint measurement of the loan quantities
and the existence of order effects, which are consequences that deserve a
deeper study.

On the other side, the loan entanglement is suitable for loan pools or
common sovereign bonds that could be virtually non-defaultable \cite{hatch}
or the design of diversified portfolios with lower correlations between the
different investments to avoid the turbulence in the financial markets \cite%
{life}. Empirical studies of stress testing for portfolios of auto loans has
been shown that loans aged five years or more have significantly higher
default probabilities, but the reliability of the stress test results are
limited by the instability of the estimated coefficient of macroeconomic
variables. The loan entanglement allows to develop sensitivity analyses and
make conservative adjustment to minimize model risk \cite{wu}. In turn, the
vast variety of markets across the world trade a broad range of financial
products, and the prices of the assets traded are sensitive to the market
news, which give a strong coupling between them \cite{fenn}. In turn,
interaction of loan diversification and market concentration indicates that
diversifying banks operating in highly concentrated markets are more
financially stable compared to those in less concentrated markets \cite{shim}%
. It has been shown that debt restructuring is significantly easier for
loans from traditional bank lenders than loans from institutional lenders 
\cite{demi}. The loan algebra introduced above can gives an unified solution
to these problems by making more flexible, not the initial conditions of the
loan, but directly the loan time evolution. In turn, the loan entanglement
can be used to develop optimal debt structures when the moral hazard problem
is severe \cite{park}. From the mathematical viewpoint, a linear algebra
implies to conceive all possible realizations of the financial quantities.
For amortization systems, the loan quantities can have any positive real
number, subject to restrictions given by the preserved distance in the
vector space, but it can be used to model secondary financial markets, where
all possible realizations of investors holding securities and cash is taken
as the basis of the Hilbert space of market states and the temporal
evolution of an isolated market is unitary in this space \cite{schaden}. In
turn, GHA is suitable to model the quantum anharmonic oscillator, which is
used as a model for the stock market \cite{tgao}, where the motion of the
stock price is modeled as the dynamics of a quantum particle or it can be
used to model supply and demand as different potentials appearing in the
Hamiltonians (\cite{orr} and \cite{berr}).

\section{Conclusions}

In this work we have enlarge the benefits of rewriting the amortization
systems on vectorial spaces. By a suitable choice of an algebra of operators
for the debt, amortization, interest and periodic installments that act on a 
$M$-dimensional vectorial space, where $M$ is the loan duration, the usual
recurrence relations for the amortization systems are found in terms of the
eigenvalues of these operators. Given the $SO(M)$ symmetry of the vectorial
space, a basis rotation of the orthonormal basis allows us to change the
schedule of the periodic payments, allowing better benefits for the borrower
in the case a payment cannot be afford and the borrower can be classified as
a defaulter. Superposition of the classical amortization systems such as
French and German systems are studied showing the possibility of creating
new amortization systems that combine constant amortizations and periodic
payments. In turn, for indexed loans, where the debt and the payments are
linked to macroeconomical indices such as inflation, the rotation allows the
borrower to avoid the increment of the payments by selecting specific angles
of rotation. Using the tensor product of vectorial spaces, different loans
can be entangled with procedure analogous of the quantum prisoner's dilemma.
By introducing an entanglement operator and allowing each borrower to apply
its own rotation of the vectorial space, the payment schedules gets
entangled through an entanglement parameter, which can be defined prior to
the beginning of the repayment by the lenders or borrowers of both. The
results obtained are a generalization of the classical amortization systems
and can be conceived as a new financial instrument for debt repayment of
private entities or sovereign countries.

\section{Acknowledgments}

This paper was partially supported by grants of CONICET (Argentina National
Research Council) and Universidad Nacional del Sur (UNS) and by ANPCyT
through PICT 2019-03491 Res. No. 015/2021 and PIP-CONICET 2021-2023 Grant
No.11220200100941CO. J. S. A. is a member of CONICET.

\section{Data availability}

No datasets were generated or analyzed during the current study.

\bigskip

\end{document}